# Cloud-Based Secure Authentication (CSA) Protocol Suite for Defense against DoS Attacks

Marwan Darwish[1], Abdelkader Ouda, Luiz Fernando Capretz

Department of Electrical and Computer Engineering

The University of Western Ontario

London, Canada

{mdarwis3, aouda, lcapretz}@uwo.ca

**Abstract** — Cloud-based services have become part of our day-to-day software solutions. The identity authentication process is considered to be the main gateway to these services. As such, these gates have become increasingly susceptible to aggressive attackers, who may use Denial of Service (DoS) attacks to close these gates permanently. There are a number of authentication protocols that are strong enough to verify identities and protect traditional networked applications. However, these authentication protocols may themselves introduce DoS risks when used in cloud-based applications. This risk introduction is due to the utilization of a heavy verification process that may consume the cloud's resources and disable the application service. In this work, we propose a novel cloud-based authentication protocol suite that not only is aware of the internal DoS threats but is also capable of defending against external DoS attackers. The proposed solution uses a multilevel adaptive technique to dictate the efforts of the protocol participants. This technique is capable of identifying a legitimate user's requests and placing them at the front of the authentication process queue. The authentication process was designed in such a way that the cloud-based servers become footprint-free and completely aware of the risks of any DoS attack.

## 1. Introduction and related works

Cloud computing is the utilization of hardware and software to provide services to end users over a network, such as the Internet. Cloud computing includes a set of virtual machines that simulate physical computers and provide services, such as operating systems and applications. However, configuring the virtualization within a cloud computing environment is critical when deploying a cloud computing system. A cloud computing structure relies on three service layers: Infrastructure as a Service (IaaS), Platform as a Service (PaaS), and Software as a Service (SaaS) (Figure 1). IaaS provides users with access to physical resources, networks, bandwidth, and storage. PaaS builds on IaaS and provides end users access to the operating systems and platforms necessary to build and develop applications, such as databases. SaaS provides end users with access to software applications.

---

[1] Corresponding author. Tel.: +1-226-688-8214
E-mail address: mdarwis3@uwo.ca
Mail address: Department of Electrical and Computer Engineering
Thompson Engineering Building, Western University
London, Ontario, Canada, N6A 5B9

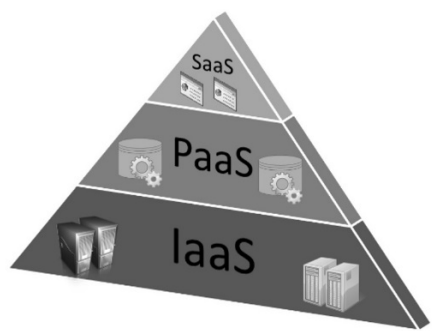

Figure 1. Cloud computing architecture

DoS attacks represent major security risks in a cloud computing environment, where the resources are shared by many users. A DoS attack targets the resources or services in an attempt to render them unavailable by flooding the system resources with heavy amounts of artificial traffic. Dealing with DoS attacks at all layers of cloud systems is a major challenge due to the difficulty of distinguishing the attackers' requests from legitimate user requests, particularly when the data are transferred between the layers of the cloud computing systems. Therefore, detecting a DoS attack in its early stage, in the upper layer (SaaS), is a significant approach to avoid the destruction caused by DoS attacks on the other layers. However, all service requests for SaaS must be authenticated to be approved.

Many authentication protocols can be used in the SaaS layer. The OAuth protocol [1] is currently a widely used authentication protocol that controls the access of third-party applications to a HTTP service. In OAuth, the resource owner can allow a third-party client to access the resources through the owner. For example, a user as a photo owner (resource owner) can grant permission to a printing service (client) to access the user's photos. The photos are stored on photo exchange server as a (resource server). Rather than sharing the user's credential with the printing service, the user is authenticated to a server that is trusted by the photo exchange server (authorization server), which then issues a credential (such as an access token) to access the resources. There are limitations when the owner shares credentials, such as a username, and password with the third-party client to access restricted resources. The first limitation is that the access information includes the password, which is most likely stored by a third-party client as clear-text for future access. The second limitation is that the server should only use a password as an authentication method. The third limitation is that the resource's owner cannot limit the access of a third-party client and also cannot control the duration of the access. Finally, if the password is accessible in a third-party client, all the resources will be accessible, as well. Therefore, OAuth allows a third-party client to access the resources of the server with privileges and rules without using the resource owners' access information. The process of the protocol, as shown in Figure 2, operates in the following manner:

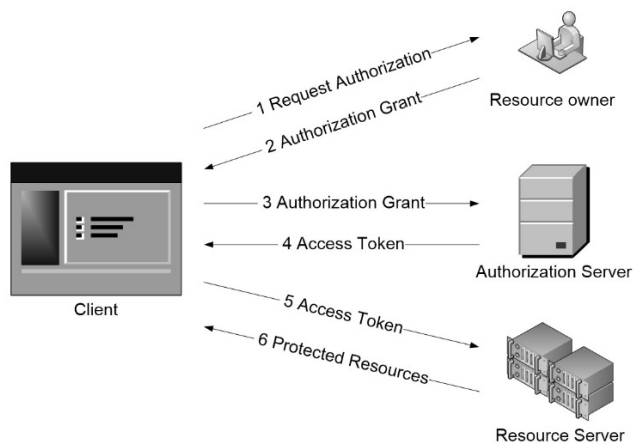

Figure 2. The process of the OAuth protocol.

(1) The protocol process starts when the resource owner receives an authorization request from the client.

(2) The resource owner sends back the authorization grant to the client. The clients' authorization request determines the type of grant. Examples of the different types of grant are as follows:

- **Authorization code grants** that are given to the client by the resource owner after the resource owner has been authorized by the authorization server. With such a code, the client does not require the resource owner credentials, and it is a secure grant type.
- **Implicit grants** that depend on browser implementation using a scripting language, such that the access token is issued directly to the client. The implicit grant minimizes the flow process of the protocol but also leads to security issues.
- **Resource owner password credentials grant** that uses the resource owner's credentials (username and password) to issue an access token to the client. This type of grant can be used when the client is highly trusted by the resource owner.
- **Client credentials grant** that can be used for a limited scope of access to protected resources on the server. This type of grant is used when the client is a resource owner or when the client previously had privileges to access the protected resource.

(3) The client sends an authorization grant and the authentication to an authorization server to obtain the access token.

(4) The access token will be provided to the client once the client is authenticated and the authorization grant is validated by the authorization server. In this case, the access token replaces the typical authentication, such as username and password, and is also recognized by the resource server. The access token can be used in different methods based on the security requirements of the server, such as with different types of cryptography. When the token is expired or became invalid, the token can be refreshed as an optional process by sending the authorization grant to an authorization server, whereupon the client receives a request with the access token and refresh token.

(5) The client sends the access token to request restricted resources from the resource server.

(6) The server will respond to the request when validating the access token.

However, any insecure implementation of OAuth protocol can lead to the possibility of a DoS attack.

Many authentication protocols have been proposed for the SaaS layer, but they are unaware of a DoS attack. Yassin et al. [2] proposed an authentication process based on a one-time password (OTP) with mutual authentication of the user and the cloud server. Yassin's authentication schemas defended against the possibility of a replay attack, but not against a DoS attack. Some cloud-based authentication protocols for DoS prevention have been proposed, such as those by Choudhury et al. [3], Hwang et al. [4], Jaidhar [5], and Tsaur et al. [6], but they use a smart card reader for the authentication process. Furthermore, the Yassin et al. [7] schema also recommend the use of an extra physical device, such as a fingerprint scanner.

On their own, the authentication protocols can lead to vulnerability to a DoS attack. Therefore, it is necessary and significant to verify the DoS-resistance in every process of the authentication protocol. For example, on the one hand, verifying a large number of signed messages via the server consumes the resources of the server to a significant degree, particularly when the attacker sends a massive number of forged signed messages. On the other hand, sending a typical client credential with each request in the authentication protocol will force the server to verify these requests based on the stored information at the server. As a consequence, the server resources will be exhausted when dealing with a large number of requests.

An example of authentication protocols that can introduce internal DoS risks on their own is shown in Figure 3. The goal of this protocol is to authenticate both the client and the server to each other. This protocol uses the ephemeral Diffie-Hellman key-exchange [8], where a, b, p, and g are the values of Diffie-Hellman. In this protocol, once the server receives a request from a client, the server will begin generating the secret value b. Subsequently, the server will compute the exponential value, $g^b$ mod p. Moreover, the server will encrypt the *nonce of* the client and the exponential value via the client public key. Finally, the server will digitally sign the encrypted message. All of these processes will be executed by the server, which consumes a great deal of resources without determining whether the request is legitimate. This mutual authentication, which is vulnerable to DoS attack, is similar to the two-way authentication version of the Transport Layer Security (TLS) protocol [9]. As another example of a protocol that introduces DoS risk on its own is Kim et al.'s protocol [10], which aims to securely authenticate the key exchange between participants. In this protocol, once the server receives the first message, the server will begin computing an exponential value and generate the key, and as such, the server resources can become exhausted by the initial requests.

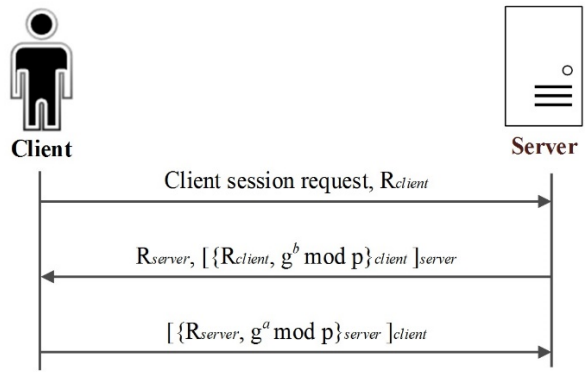

Figure 3. Mutual authentication protocol

An example of an authentication protocol that is aware of a DoS attack in the traditional network is a Host Identity Protocol (HIP) [11]. However, this protocol cannot be implemented in the application layer because it is based on the host identity on the network layers in the OSI reference model, and it is configured and controlled at an operating system level. Moreover, any authentication protocol that is based on IP address verification, such as the IPSec protocol, makes it difficult to hide the identity of the participants.

In this work, we present a novel cloud-based authentication protocol suite that can defend against external DoS attacks and also conveys an awareness of internal DoS attacks. The proposed protocol suite uses an adjustable technique to direct the efforts of the protocol participants. This technique can recognize a legitimate user's requests and passes such requests to the authentication process. The authentication process was developed to become fully aware of any possibility of DoS attacks while avoiding the use of any extra physical devices. Section 2 describes the CSA protocol against DoS attacks in the SaaS layer. Section 3 demonstrates the security of the CSA protocol against DoS attacks. Finally, section 4 briefly summarizes the work.

## 2. Cloud-based secure authentication (CSA) protocol suite

The authentication protocol must be investigated in terms of its vulnerability to DoS attack by using a cost-based model approach [12]. The cost-based model approach proposed by Meadows aims to prevent DoS attacks during the authentication process. This model depends on the exhausted resource costs of the participants. The cost-based model approach logically demonstrates the effectiveness of the protocols in preventing DoS attacks. In this approach, the *computation cost* is defined as the total resource usage cost of the requester (client) and responder (server) when both participate in the authentication protocol. The cost is computed during the process until the DoS attacker is detected and is prevented from participating. The *total cost of the requester* is the total estimated cost of each operation involved in the authentication process on requester's side until the authentication process ends. However, the *total cost of the responder* is the total estimated cost of each operation during the authentication process until the requester is determined to be either a legitimate requester or attacker.

Meadows proposed the following categories for an operation's cost: inexpensive, medium, and expensive. This approach assumed that the exponential, check signature and signature operations performed during the authentication process are expensive. The pre-calculated exponential value, encryption, and decryption operations are of medium cost. Any other operation is inexpensive. Therefore, the CSA protocol was developed so that the total resource cost of the client's side will be greater than the resource operations cost of the cloud-based server when they participate in the authentication process together. Table (1) shows the notations that are used in the CSA protocol suite.

Table 1 - Notations of the CSA protocol

| Notation | Description |
|---|---|
| *client* | The cloud user |
| *cloudserv* | The cloud server/service provider |
| CID | Client ID |
| UET | Unique encrypted text; the key of the UET is known only by *cloudserv* |
| SK | Session key |
| A | A set of random integers of the server challenge function |
| S | A subset sum of the server challenge function |
| B | A binary vector representing the challenge function solution |
| $R_{cloudserv}$ | The nonce that is generated by *cloudserv* |
| T | Timestamp |
| MK | Master secret key of *cloudserv* |

CSA consists of sets of protocols. The first protocol is used for the registration process, which is an agreement process between the participants (*client* and *cloudserv*) about certain shared information. Thus, the participants can use that information during the operation of other CSA protocols. The second protocol is an adaptive protocol that works against DoS attacks. This protocol was developed based on the cost-based model approach. The third protocol is used for the authentication process, which includes all operations that occur based on the initially agreed-upon information of the previous protocols. As a result, *cloudserv* can confirm the identity of the *client* and then complete the authentication process, or it can detect and then prevent an intruder in the case of a DoS attack.

**2.1 Registration protocol**

In the CSA registration protocol, the *client* and *cloudserv* will share the required identity data to register the *client* into the *cloudserv* database. As shown in Figure 4, the registration process begins when the *client* submits all the required information to *cloudserv*. This information involves the first name, last name, organization name, email address, or any other information that is required by the cloud service provider. *Cloudserv* will then verify the received information, store it in a database, and then it will send a validation email message to the *client* to confirm the *client*'s information. After validation, *cloudserv* will activate the *client*'s account. At the same time, *cloudserv* will generate a Unique Encrypted Text (UET) that is encrypted by the *cloudserv*'s master key (MK), which is known only by the *cloudserv*. The UET contains *client* information, such as the Client ID (CID), as well as any other information that will be created by *cloudserv* during the processes of the CSA protocols. The UET is a piece of information that will not be stored on *cloudserv*; rather, it will be sent to the requesting *client*. Once the *client* receives the required data from *cloudserv*, both *client* and *cloudserv* will agree regarding the pre-shared key. The pre-shard key will be created using a key derivation function and a shared secret. The *client* and *cloudserv* will agree upon the key derivation function and a shared secret at the end of the registration protocol, which will be exchanged via a secure channel in a very restricted environment. This approach is very much similar to the pre-shard key agreement (PSK) used in the UMTS and WPA2 protocols [13]. Consequently, the *client* will store the UET and a pre-shared key for a future authentication process.

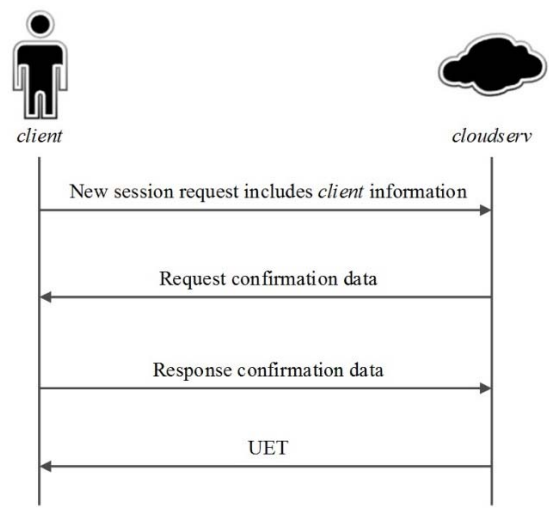

Figure 4. Registration protocol

Even if a *client* is registered to the *cloudserv*, the *client* cannot access the services available through the *cloudserv* unless *cloudserv* authenticates the *client*. To perform the authentication protocol that is ready to defend any internal or external DoS attacks, CSA provides an outer shield to the authentication protocol to help identify the legitimate clients from the DoS attackers. The CSA-adaptive DoS defender protocol is designed to provide this outer shield in a manner described in the following sub-section.

**2.2 CSA-adaptive DoS defender protocol**

The CSA-Adaptive DoS Defender Protocol utilizes the cost-based model approach that can be briefly re-stated as follows. Before applying the computational power of the authentication protocols in the server side, the clients are asked to prove their sincere commitment for receiving the cloudserv services. This validation of commitment can be achieved by any technique that can force the clients to utilize a significant amount of computational power, before the servers utilize them, to confirm their genuine requests. Currently, "Client puzzles" is a common technique that realizes the cost-based model approach [14].

In this work, a one-way function based technique is proposed to realize the cost-based model approach. A cryptographic knapsack problem was chosen, not only because it is a strong one-way function but also for its flexibility to be adaptive [15]. The strength of the knapsack problem comes from being one of those problems known to be a NP-complete problem [16]. In addition, the main characteristic of the selected one-way function in this protocol is an adjustable difficulty level for solving the puzzle based on the demanded efforts on the client's side.

In cryptography, a knapsack problem is described as follows: given a set of positive integers (i.e., items) A= $a_1$ … $a_n$ and a positive integer value S, is there a non-empty subset of $a_1$ … $a_n$ whose values sum to S? For example, let the set of items in a knapsack A be (13, 54, 28, 73, 3, 36) and the summation S be 89. Therefore, finding the elements 13, 73, and 3 solves the problem because their summation is equal to 89. In other words, finding a binary vector B such that A × B = S solves the problem. In this example B is the vector (1, 0, 0, 1, 1, 0), and hence A×B = 13 + 73 + 3 which is 89.

Typically, the complexity of the knapsack problem depends on the size of the knapsack A (the number of its items, say n) and on the number of 1s (say m) in the binary vector B. Note that if the number of items in A, i.e., n, is small, then an exhaustive search for the solution is practical. Note also that, if the number of 1s in B, i.e., m is small comparing to (n), then a solution can be found in a reasonable time via dynamic programming algorithms.

Consequently, by adjusting the values of n and m, determining the difficultly level of the knapsack problem and hence the cost-based approach can be adaptively realized. The CSA protocol considers n = 512 items. These items are fixed integer values that both parties should agree upon during the registration protocol. Based on the experimental result (see section III-B), obtaining vector B will force the *client* to become involved in finding the solution to $2^{512}$ subsets, which is a highly time and resource consuming process. The number of subsets of items is adjustable based on the required efforts of the participants. Moreover, the chosen items that are used during the summation process are determined by hashing the values of CID, MK, and $R_{cloudserv}$ using SHA2-512, where MK is the master secret key of the *cloudserv*. Note that the result of the hash function is a 512-bit stream. Moreover, the subset of the 512-bit stream that includes a specific number of ones (m) represents the required vector, B, of the knapsack problem. For example, if the protocol developed to let m = 55, the *cloudserv* will take the subset of the 512-bit stream that includes the first 55 ones; increasing the m value will make the process of solving the puzzle harder while also increasing the time-consuming nature of the puzzle-solving process. The hashing process is mandatory to verify the subset summation value (S) of the *client* after the calculation process.

The adaptive DoS defender protocol process shown in Figure 5 functions as follows:

1) *Client* sends a request for a service with CID to *cloudserv*.

   At this point, *cloudserv* will block any CID that has performed three consecutive requests within a low time threshold to prevent DoS attacks. The attacker may attempts to launch a DoS attack by sending requests with randomly generated CID values. In this case, the *cloudserv* resources will be less affected than when checking each request for information from a database system because *cloudserv* will simply reply to each request with an S value.

2) *Cloudserv* will reply directly to the *client* by sending the puzzle element as a challenge, which is the subset summation value (S) along with a *cloudserv* nonce, $R_{cloudserv}$, to the *client*.

   *Cloudserv* will ask the *client* to prove its sincere commitment for receiving the *cloudserv* services by asking for the UET as well as the puzzle solution to the (S) value. The expected solution for this challenge is the vector B.

3) Once the *client* performs a calculation and obtains vector B, the *client* will send the UET, vector B, the value of S, the received $R_{cloudserv}$, CID and the encrypted timestamp T to *cloudserv* for validation. Note that the notation $E(T, K_{pre-shared})$ means that the timestamp T is encrypted by the pre-shared key K.

   At this point, *cloudserv* has all the information required to validate the authentication requests, so *cloudserv* can apply the validation process to only a few operations, such as the following:

   - *Cloudserv* will check the subset of item $a_i$ by securely hashing (CID, MK, $R_{cloudserv}$) and comparing the result vector with the received vector B to determine whether they are similar.
   - *Cloudserv* will check the time difference between the received encrypted timestamp T and the current time stamp to determine whether it is a reasonable time difference in which to find the solution.

If any of the two previous conditions do not apply, *cloudserv* will drop the request and consider it to be an attacker's request. However, once the *client* request passes the two conditions, *cloudserv* will decrypt the UET and validate the decrypted information that contains the CID.

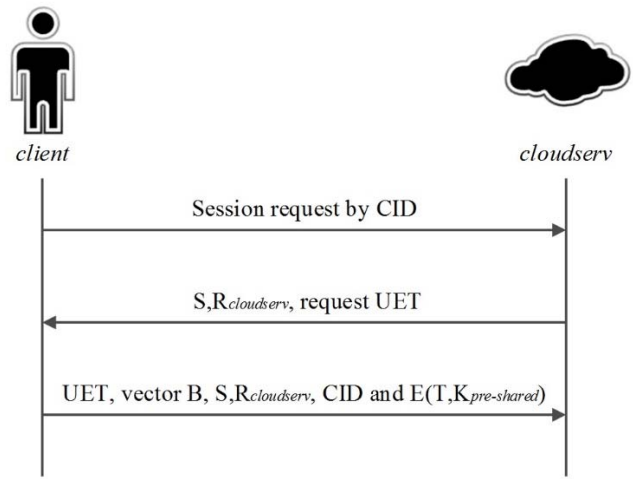

Figure 5. Adaptive DoS defender protocol

To complete the authentication process after the adaptive protocol determined the *client* as being legitimate, the CSA protocol develops an authentication protocol. The participants in the authentication protocol will agree on the session key for future interactions. In addition, they can agree on the sub-session key if they require a refreshment process later.

**2.3 Authentication protocol**

After the validation process in the previous protocol, *cloudserv* will generate the Session Key (SK), which is encrypted via a pre-shared key. Moreover, *cloudserv* will add both the SK and T information to the UET. Consequently, *cloudserv* is protected against DoS attacks to the storage space because UET will never be saved in the *cloudserv*. Furthermore, *cloudserv* can apply the refreshment property of the session key for future communication by adding the SK to the UET.

Therefore, the authentication protocol, as shown in Figure 6, performs as follows:

1) *Cloudserv* will send to the *client* the generated SK that is encrypted by the pre-shared key, along with the modified UET.
2) *Client* will confirm the received encrypted SK by sending back the modified UET and the encrypted timestamp T to the *cloudserv*. Therefore, *cloudserv* will decrypt the UET, validate the CID and obtain the SK, then confirm it by decrypting the received timestamp T using the SK.

Later, the two parties can agree regarding the sub-session keys by re-applying the processes of the authentication protocol so that the *cloudserv* can generate a sub-session key and add it to the UET without storing it in the cloud system.

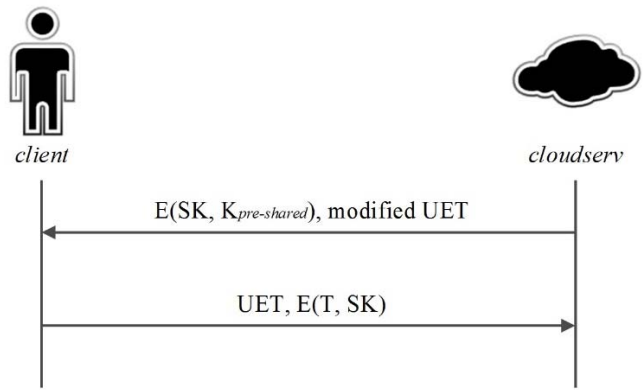

Figure 6. Authentication protocol

## 3. Analysis of the CSA protocol suite

Assessment of the CSA protocol entails evaluation of the protocol's efficiency against DoS attacks by applying a cost-based model approach. In addition, the evaluation process measures the requester's time spent when the *client* participates in the puzzle-solving process during the authentication process.

**3.1 Validation of the CSA protocol suite via a cost-based model approach**

As shown in Table II, based on the cost-based model approach, the operation cost of the *client* for the CSA protocol is categorized as expensive, particularly when the *client* solves the puzzle. The other operations of the *client* are listed within the medium or inexpensive categories. However, the maximum operation costs of *cloudserv*, including the pre-calculation and decryption operations, are within the medium category. As a result, the CSA protocol suite is an effective protocol against DoS attacks, in which the consumption cost for the requester is higher than the consumption cost for the cloud service provider during the authentication process.

Table 2 - Validation of the CSA protocol suite via a cost-based model approach

| *Client* | | *Cloudserv* | |
|---|---|---|---|
| **Operation** | **Cost Category** | **Operation** | **Cost Category** |
| Send the initial request. | Inexpensive | Reply directly to the request via secure hashing of the received values to obtain the puzzle element and ask the *client* for the UET. | Inexpensive-Medium |
| Solve the puzzle until the result is obtained. Then, send the result and the UET to *cloudserv*. | Expensive | Verify the received elements. | Medium |
| Decrypt the session key (this operation occurs after the prevention of possible DoS requests). | Medium | Decrypt the UET. Generate and encrypt the session key (this operation occurs after the prevention of possible DoS requests). | Medium |

**3.2 Analysis of the response time for the puzzle-solving process**

During the registration protocol, both parties acknowledge the fixed number of items n that can be used in the puzzle-solving process. Consequently, it is important to identify the number of combination items that leads to an acceptable response-time on the requester's side during this process. An experiment that uses various numbers of combination items is performed to estimate the average response-time in seconds for each combination. In our experiment, we used a dynamic programming algorithm to solve a subset sum puzzle problem of n =100 items. The experiment was performed using a i7-4770 CPU at 3.4GHz with 32 GB RAM. Our experiment indicates that applying dynamic programming to n = 100 items with m = 80 combination items can solve the subset sum problem in approximately 8 seconds, which is still a small amount of time to be consuming the requester's resources. Tritilanunt et al. [17] implemented the $L^3$ algorithm developed by Lenstra et. al. [18] to solve the subset sum problem. They limited their experiment to n = 100 items due to the memory-exhausted limitation. Moreover, their experiment indicated that the $L^3$ algorithm can solve the subset sum problem of n = 100 items with m = 80 combination items in 2.7 k seconds. While our experiment with the same number of items n and the same combinations of items m indicates that the subset sum problem can be solved in approximately 8 seconds. Therefore, depending on the $L^3$ algorithm to develop an authentication protocol with a subset sum problem of n = 100 items makes the protocol itself vulnerable to a DoS attack. As a result, we applied dynamic programing to a subset sum problem of n = 512 items. We experimentally found that m between 50 to 60 combination items causes approximately 20 seconds of delay on the *client's* side, as shown in Figure 7. Based on a study performed by Nielsen [19], 20 seconds is a reasonable response time that is affected by the process of solving the subset sum puzzle problem, such that the legitimate user's resources turn out to be busier for a period of time with each initial request. At the same time, this resource-exhausting process will influence an attacker who launches a DoS attack with a massive number of requests from his device. If the attacker uses many devices to launch DoS attacks, the cloud system will not be exhausted because the attack will be detected at the early stage of the authentication process.

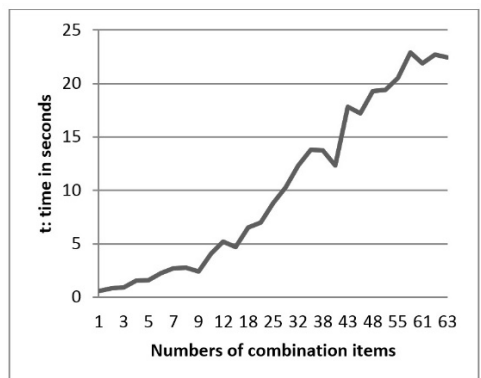

Figure 7. The response time of the requester (in seconds) with various number of combination items

## 4. Conclusions

The use of software systems in a cloud-computing environment is increasingly common. Verifying users via an authentication protocol is considered to be an initial stage to access these systems. Consequently, the authentication protocol is a main target of attackers implementing a DoS attack that decreases the availability of cloud services. Using existing strong authentication protocols of traditional network systems in cloud-based applications may lead to DoS attack vulnerability because the initiation of a massive amount of authentication processes could exhaust the cloud's resources and render the cloud-based application unreachable. In this study, the proposed CSA protocol suite aimed to prevent internal and external risks to DoS attacks. The CSA protocol uses an adaptive challenge technique based on the required efforts of the participants. Using this technique allows the system to identify legitimate requests and pass them to the cloud applications. This CSA protocol suite does not require any external physical device for the authentication process. The effectiveness of the CSA protocol was analyzed in this work using a cost-based model approach, and the ability of the protocol to fortify against a DoS attack was demonstrated.

## Acknowledgements

This work was partially supported by King Abdulaziz University through the Cultural Bureau of Saudi Arabia in Canada. This support is greatly appreciated.